%% file: main.tex
\documentclass[manuscript, nonacm]{acmart}

\AtBeginDocument{%
  }
\usepackage{makecell}
\usepackage{booktabs} % For formal tables
\usepackage{amsmath}
\usepackage{graphicx}
\usepackage{multirow}
\usepackage[caption=false, font=footnotesize]{subfig}
\usepackage{xspace}
\acmConference[Compound AI Systems Workshop]{Make sure to enter the correct
  conference title from your rights confirmation emai}{June 13,
  2024}{San Francisco, CA}
\newcommand{\stitle}[1]{\vspace{0.2em}\noindent\textbf{#1}}%%
\newcommand{\hide}[1]{}%%
\newcommand{\eg}{{\itshape e.g.}, }%%
\newcommand{\task}[1]{}%%%
\begin{document}

\title{A Blueprint Architecture of Compound AI Systems for Enterprise}


% \authornote{Equal author contribution. Authors ordered by random selection.}
% \affiliation{%

\author{Eser Kandogan, Sajjadur Rahman, Nikita Bhutani, Dan Zhang,  Rafael Li Chen, Kushan Mitra, Sairam Gurajada, Pouya Pezeshkpour, Hayate Iso, Yanlin Feng, Hannah Kim, Chen Shen, Jin Wang, Estevam Hruschka
}
\affiliation{%
  \institution{Megagon Labs}
  \country{USA}
  }
% \email{{eser,sajjadur,nikita,dan_z,rafael,kushan,sairam,pouya,hayate,yanlin,hannah,chen_s,jin,estevam}@megagon.ai}

% \affiliation{%

% \affiliation{%

% \affiliation{%

% \affiliation{%

% \affiliation{%

\renewcommand{\shortauthors}{Kandogan et al.}

\begin{abstract}
Large Language Models (LLMs) have showcased remarkable capabilities surpassing conventional NLP challenges, creating opportunities for use in production use-cases. Towards this goal, there is a notable shift to building compound AI systems, wherein LLMs are integrated into an expansive software infrastructure with a multitude of components like models, retrievers, databases and tools. In this paper, we introduce a blueprint architecture for compound AI systems to operate in enterprise settings, in a cost-effective and feasible manner. Our proposed architecture aims for seamless integration with existing compute and data infrastructure, with `stream' serving as the key orchestration concept to coordinate data and instructions among agents and other components. Task and data planners, respectively breakdown, map, and optimize tasks and data to available agents and data sources defined in respective registries, given production constraints such as accuracy and latency.

% LLMs have demonstrated impressive capabilities beyond traditional NLP problems, paving the way for productionalization. Towards this goal, we see a significant shift towards compound AI systems, where LLMs are part of a large software infrastructure, with a multitude of supporting components. In this paper we propose a blueprint architecture for enterprise to facilitate deployment of compound AI systems, in a cost-effective and feasible manner, seamlessly integrating with existing compute and data infrastructure through appropriate touchpoints and interfaces, defining existing  models, APIs, and data. To ensure effective operation of a compound AI system, we envision `stream' as the key orchestration concept that coordinates data and instructions among various components. We also introduce the concept of task and data planners for efficient execution of instructions and corresponding data operations by optimizing for production setting-specific objectives such as accuracy and latency. 

%Key orchestration concept in the architecture is `stream' as the primary means to coordinate data and instructions among agents. Furthermore, components such as task and data planners can optimize for cost and quality to accomplish tasks and retrieve data. 

% [Figures: 
% \href{https://docs.google.com/presentation/d/1UhT9lG5fG4PuPI1PUA19zeNnJHQ7WF2J9KNy20JkxjM/edit#slide=id.p}{Figures}
% ]
\end{abstract}

\maketitle
\input{intro}

\input{blueprint}
\input{conclusion}

% \begin{acks}
% \end{acks}

\bibliographystyle{ACM-Reference-Format}
\bibliography{paper}


\end{document}

%% file: intro.tex
\section{Introduction}
\label{sec:intro}

LLMs have demonstrated impressive capabilities  in various tasks that extend beyond traditional NLP problems ~\cite{petroni2021kilt,izacard2022atlas,li2022competition,schick2024toolformer,zhang2023multimodal}
%~\cite{petroni2021kilt,mishra2023characterizing,wu-etal-2024-less,zhang24coling,maekawa2024retrieval}
, ushering a new era of LLM-powered applications that leverage their abilities across multiple domains. In current approaches, LLMs assume a central role in nearly every aspect, encompassing task planning, data discovery and retrieval, and interfacing with other tools and services. However, such extensive involvement often poses challenges to deployment in production settings, where additional task and data constraints, such as latency, accuracy, cost, availability and quality, among others, must be considered ~\cite{xi2023rise,yang2024harnessing,wu2023autogen,khattab2023dspy}.
% ~\cite{kim2023llm, li2023camel, wu2023autogen,}.

% However, this comprehensive involvement often poses challenges to deployment in production settings ~\cite{kim2023llm, li2023camel, wu2023autogen}. Within such a setting, there additional constraints beyond simple tasks execution ---  such as latency, accuracy, and cost, among others --- that necessitate instrumentation of optimization strategies.

Towards productionalization, there is a shift from monolithic models to compound AI systems that incorporate various components other than LLMs, e.g. components for data retrieval, control flow,  proprietary models, and databases. Such systems provide enhanced performance for complex tasks, greater flexibility and adaptability across different use cases, easier integration of existing models and data, and greater control and trust. For example, LinkedIn and Indeed, two global job matching and hiring platforms, are productionalizing compound AI systems for a multitude of tasks in HR such as matching, recruitment, and career guidance, among others~\cite{indeed-ai-blog, linkedin-ai-blog}.

% Towards the goal of productionalization, we see a significant shift towards compound AI systems~\cite{compound-ai-blog}, where LLMs still play an important role but they are part of a large software infrastructure, with a multitude of components (agents and beyond), to plan and break-down complex tasks, to discover and query proprietary data, and to exploit proprietary models and services, and an underlying system that orchestrates the flow of data and control among components to seamlessly function together. 

We propose a blueprint architecture of a compound AI system tailored for enterprise use unlike existing work~\cite{wu2023autogen, Liu_LlamaIndex_2022} which lack support for agent orchestration and optimization of agentic workflows. Key factors we consider in the design include:  (1) ensuring \emph{seamless integration} into existing infrastructure through suitable touch points and interfaces, and (2) effectively \emph{orchestrating} work within and external to the compound system with appropriate resource allocation, and (3) maximizing \emph{utilization} of the system in a cost-effective manner.

% In this paper we propose a blueprint architecture for enterprise to facilitate the use of LLMs in production in a cost-effective and feasible manner fitting into the existing infrastructure. Key concept in the architecture is `streams', as the primary means to orchestrate data and instructions among agents. The architecture defines important touch points and interfaces to allow seamless integration with existing compute and data infrastructure, where existing deployed models, APIs, and data can be utilized. Furthermore, key components in the architecture such as task and data planners can optimize for cost and quality to accomplish tasks and retrieve data. 

%% file: blueprint.tex
\section{Blueprint Architecture}
\label{sec:overview}

\begin{figure}[!htb] 
  \vspace{-10pt}
  \centering
  \includegraphics[width=\linewidth]{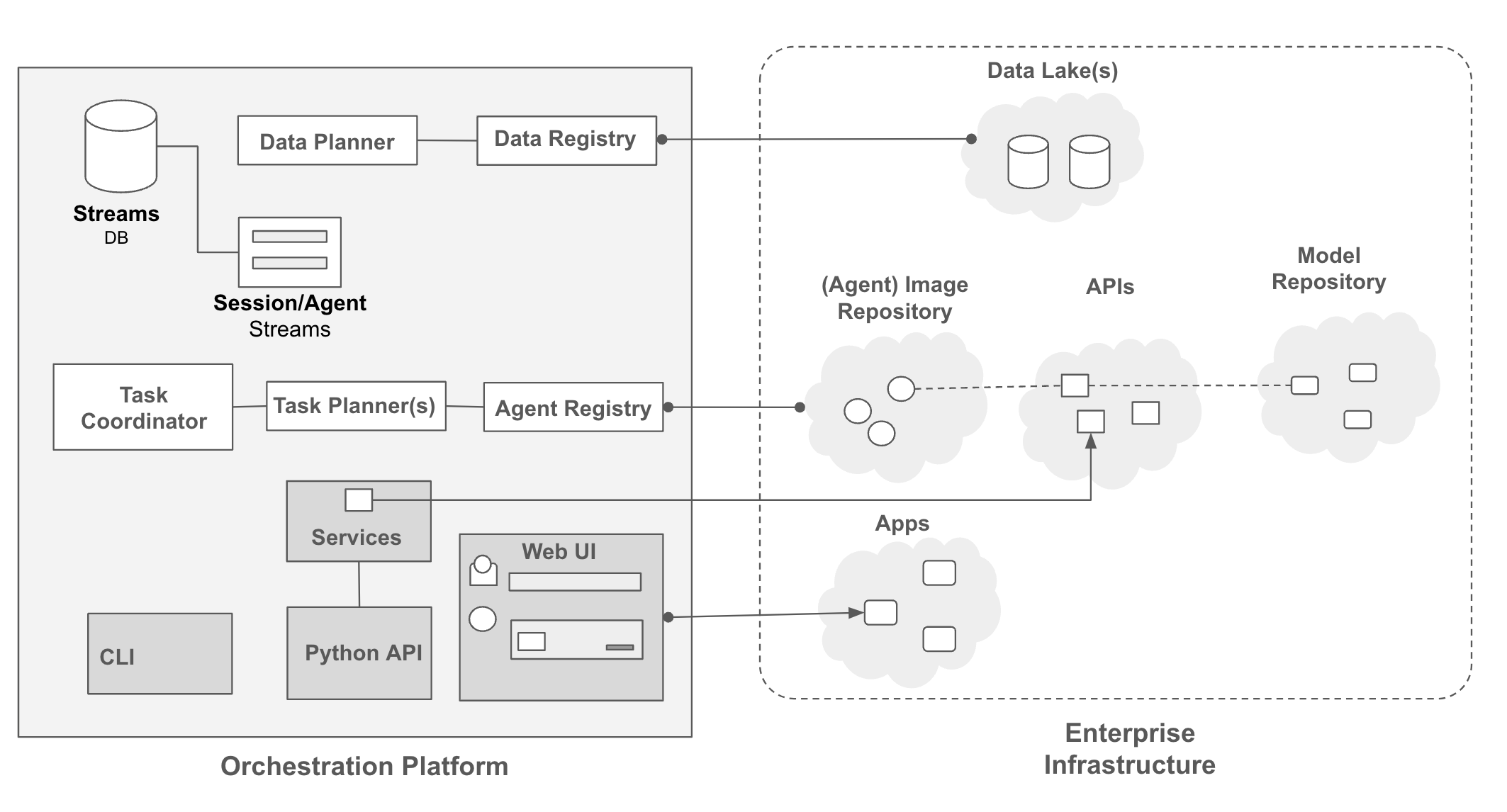}
  \caption{Blueprint Architecture: Data and Agent Registries are touch points that define existing data, models, APIs, and services in the enterprise for utilization by agents.}
  \label{fig:architecture} 
  \vspace{-15pt}
\end{figure}

Key components in the blueprint architecture include: (1) \emph{agents, agent and data registries} as key touch points and interfaces to seamlessly integrate with existing deployed models, APIs, databases, and tools, (2) \emph{streams} to orchestrates data and instructions across components, and (3) \emph{task and data planners} to optimize for cost and quality constraints in task execution and data retrieval (Figure~\ref{fig:architecture}).

%We next describe the key components of the architecture towards these considerations (Figure~\ref{fig:architecture}).

% Key considerations for designing a blueprint architecture for compound AI systems for production are: (1) \emph{utilization} of overall system in a cost effective manner, (2) seamless \emph{integration} into existing infrastructure through appropriate touch points and interfaces, and (3) \emph{orchestration} of work within and external to compound AI system with appropriate allocation of resources.

\subsection{Integration: Touchpoints and Interfaces}

\begin{figure}[!htb] 
  \centering
  \includegraphics[width=\linewidth]{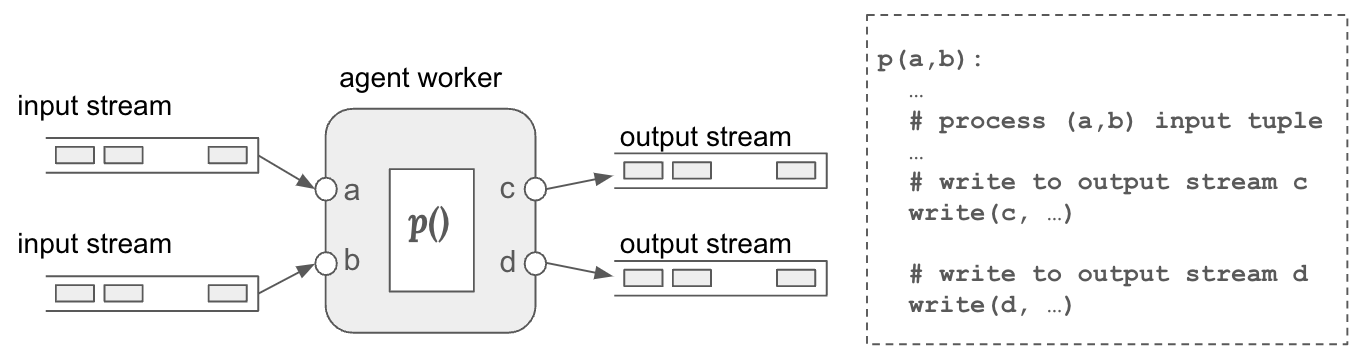}
  \caption{Agents: Triggered by data/instruction messages from multiple incoming streams agents process and produce output data and instructions to multiple output streams.}
  \label{fig:agent} 
    \vspace{-10pt}
\end{figure}

\stitle{Agents.} Agents are `compute' constructs to perform tasks (Figure~\ref{fig:agent}). They do so by calling service APIs (e.g. JobSearch), interfacing with LLMs (e.g. OpenAI), running predictive models (e.g. MatchPredict), etc. As part of agent specification, inclusion and exclusion rules dictate when agent execution gets triggered.  To improve utilization and concurrency, each agent has a group of workers, which process input data through a `processor' function, defined by the agent.

\stitle{Agent Registry.}
% An agent registry serves as a metadata of agents, with capabilities to search agent metadata (e.g. agent descriptions, inputs, outputs). 
An agent registry stores and organizes metadata (e.g., agent descriptions, inputs, outputs and specifications) of agents, and supports search and retrieve functions. Existing APIs, tools and models in the infrastructure can be defined as agents, with their descriptive metadata in the registry.

\stitle{Data Registry.} Analogous to an agent registry, data registry stores metadata of data in the enterprise, and as such is a key touch point to the existing data infrastructure. Data registry is key to aiding in the search and discovery of multi-modal enterprise data stored in data lakes and warehouses at various levels of granularity (e.g., raw data, summaries, and metadata such as schema). For instance, a lake-house architecture 
%\cite{armbrust2021lakehouse} 
can facilitate agents to operate over both data lakes (e.g., models) and data warehouses (e.g., OLAP).

\subsection{Orchestration: Streams and Sessions}

\stitle{Streams.}
A `stream' is the central `orchestration' concept in the blueprint architecture. A stream is essentially a sequence of messages, \eg data, instructions, that can be dynamically produced, monitored, and consumed. They serve as the universal communication facilitators. For instance, a user typing text in a chat can be modeled as a stream, with each word as individual messages. Similarly, an LLM agent generating content can be another stream. Streams can contain data and instruction messages and can include data of various types, \eg int, str, json. 

\stitle{Session}. `Session' is the key `context' concept that defines the scope of work, where agents join and accomplish the overall task.

\begin{figure}[!htb] 
  \centering
  \includegraphics[width=\linewidth]{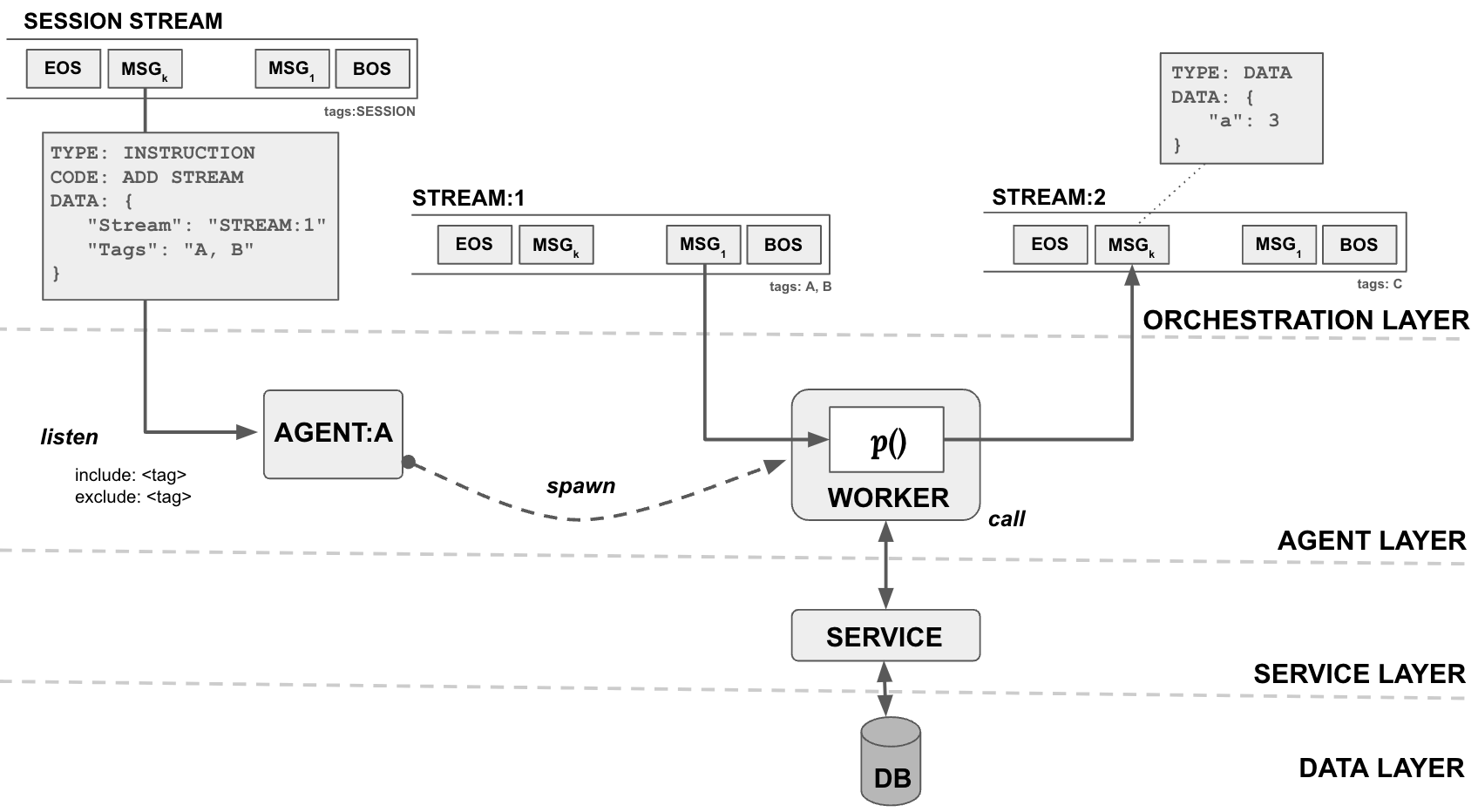}
  \caption{Orchestration of Agents: As agents join a session and generate output streams, these events are broadcasted in session streams. Other agent may choose to respond to a stream, and initiate a worker to process data in the stream, interacting with external services and databases. Computation occurs in various layers for optimal utilization. }
  \label{fig:layers} 
    \vspace{-10pt}
\end{figure}

Streams and sessions together facilitate an event-driven orchestration as shown in Figure~\ref{fig:layers}. A user agent initializes a session by creating an initial stream. Other agents are then added either by the user (or by default as part of session configuration) to the session to coordinate a response to the initial user input. Each agent announces when it joins and leaves the session in the session stream. Agents have the capability to behave autonomously by listening to a stream. 
% Agents can monitor the session stream and decide to listen to streams produced by other agents. 
If they decide to listen, they process data in the input stream and generate data (or instructions) into a new output streams within the session. Streams and messages are tagged to enable other agents to selectively consume them. Alternatively, agents can be invoked by centralized planners discussed next.

\subsection{Utilization: Planners and Coordinators}

While the architecture enables agents to accomplish tasks in a decentralized manner by utilizing stream and/or message tags, we introduce task and data planners to optimize the execution of tasks and data operations according to production constraints.

\stitle{Task Planner and Coordinator.} 
A task planner, modeled as an agent, listens to initial user agent stream and generates a task plan in the form of directed acyclic graphs (DAGs), utilizing metadata from the agent registry to identify the appropriate agents. The task planner can be interactive and adaptive, learning from user or agent feedback, in the current session, and previous sessions. Output of the task planner is also a stream containing the DAG. A task coordinator takes the DAG plan and invokes corresponding agents by issuing instruction messages with input parameters into its own output stream (which agents in the session listen to). Coordinators closely monitor and guide the execution according to the plan DAG with constraints: collecting agent output and passing it on to the following agent upon task completion (as instruction messages); intervening upon constraint violation (e.g., timeout or low-quality result); and invoking the task planner to replan when necessary.

%\sairam{Sairam: It would be nice to mention about DSPy; how a plan from task planner can be programmatically modeled using DSPy framework. Similary for a plan from Data Planner.}
%%Eser: Added to Intro.

%A task coordinator starts with the DAG plan and monitors the progress of execution. Coordinators strictly adhere to plans and activate agents by issuing instruction messages as tasks are completed, if necessary, invoking the task planner to replan. 

\stitle{Data Planner.} The data planner helps optimize the data operations within specified constraints on cost, performance and/or quality. It decomposes a complex data retrieval task into sub-tasks (e.g. discover, query, extract, summarize, join, compare). For each sub-task, it utilizes metadata from the data registry to determine the most efficient and effective way to accomplish the task.

%% file: conclusion.tex
\section{Conclusion}
\label{sec:conclusion}

We echo the sentiment by Zaharia et al.~\cite{compound-ai-blog} that a systems approach offers a viable path to develop reliable, effective and usable AI applications. While the best practices for developing AI systems is an open problem, we believe our proposed blueprint architecture will help the design and experimentation and encourage interdisciplinary research in AI, NLP, Databases, Systems, and HCI.